\documentclass[twocolumn,superscriptaddress,showpacs,prl]{revtex4-1}

\usepackage[dvips]{graphicx}
\usepackage{dcolumn}
\usepackage{bm}
\usepackage{color}

\begin{document}

\title{Coupling of magnetic and ferroelectric hysteresis by a multi-component magnetic structure in Mn$_{2}$GeO$_{4}$}

\author{J.\,S.\,White,$^{1,2}$ T.\,~Honda,$^{3}$ K. Kimura,$^{3}$ T.\,~Kimura,$^{3}$ Ch.\,Niedermayer,$^{1}$ O.\,Zaharko,$^{1}$ A.\,Poole,$^{1}$ B.\,Roessli,$^{1}$ and M.\,Kenzelmann$^4$}

\affiliation{
Laboratory for Neutron Scattering, Paul Scherrer Institut, CH~5232 Villigen, Switzerland\\
$^{2}$ Laboratory for Quantum Magnetism, Ecole Polytechnique F\'{e}d\'{e}rale de Lausanne, CH~1015 Lausanne, Switzerland\\
$^{3}$ Graduate School of Engineering Science, Osaka University, Toyonaka, Osaka 560-8531, Japan\\
$^{4}$ Laboratory for Developments and Methods, Paul Scherrer Institut, CH~5232 Villigen, Switzerland}
\date{\today}

\begin{abstract}
The olivine compound Mn$_{2}$GeO$_{4}$ is shown to feature both a ferroelectric polarization and a ferromagnetic magnetization that are directly coupled and point along the same direction. We show that a spin spiral generates ferroelectricity (FE), and a canted commensurate order leads to weak ferromagnetism (FM). Symmetry suggests that the direct coupling between the FM and FE is mediated by Dzyaloshinskii-Moriya interactions that exist only in the ferroelectric phase, controlling both the sense of the spiral rotation and the canting of the commensurate structure. Our study demonstrates how multi-component magnetic structures found in magnetically-frustrated materials like Mn$_{2}$GeO$_{4}$ provide a new route towards functional materials that exhibit coupled FM and FE.
\end{abstract}


\maketitle
The family of multiferroic (MF) materials for which ferroelectricity is induced by magnetic order continue to be of sustained interest because these materials provide genuine prospects for new and precise magnetoelectric (ME) functional devices \citep{Kim03,Che07,Ram07,Kit10}. The most promising materials are those that exhibit the coupled ferroic orders of ferromagnetism and ferroelectricity. In spinel CoCr$_{2}$O$_{4}$ \citep{Yam06}, spontaneous magnetization, $M$, and electric polarization, $P$, orders emerge when the system undergoes a transition into a transverse-conical magnetically ordered phase with a ferromagnetic (FM) component parallel to the cone axis. In materials showing proper-screw or longitudinal-conical ordered state (e.g. spinel ZnCr$_{2}$Se$_{4}$ \cite{Mur08} and hexaferrites \cite{Ish08}), the application of magnetic fields induces a transverse-conical component. Phenomenologically the generation of $P$ by non-collinear magnetic order is described by $\textbf{\textrm{P}}_{ij}\propto\textbf{\textrm{e}}_{ij}\times(\textbf{\textrm{S}}_{i} \times \textbf{\textrm{S}}_{j})$, where $\textbf{\textrm{e}}_{ij}$ is a unit vector connecting spins at sites $i$ and $j$. Microscopic descriptions consistent with this phenomenology are provided by the inverse Dzyaloshinskii-Moriya (DM)~\citep{Ser06,Moc10} or spin-current~\citep{Kat05} mechanisms. However, since $P$ is generated by a non-collinear spin order the directions of $M$ and $P$ cannot be parallel in the aforementioned materials~\citep{Yam06,Ish08,Mur08}, consistent with experimental observation.\par

In contrast, materials for which $M\parallel P$ are comparatively few. In $Re$FeO$_{3}$ ($Re=$Dy,Gd)~\citep{Tok08,Tok09}, it is found that $M\parallel P\parallel \textbf{c}$, though only for GdFeO$_{3}$ are these two orders spontaneous. In these systems an exchange striction acting between $Re$-Fe$^{3+}$ spins is able to explain the microscopic origin of $P$. Here, we present a study of Mn$_{2}$GeO$_{4}$, which also exhibits spontaneous $M$ and $P$ vectors that are both parallel to the \textbf{c}-axis. In contrast to $Re$FeO$_{3}$, the $M$ and $P$ orders arise due to different components of a single magnetic structure and, in particular, Mn$_{2}$GeO$_{4}$ is the first material for which $M\parallel P$, where $P$ can be understood as generated by non-collinear spin order~\citep{Ser06,Moc10,Kat05}.\par

\begin{figure}
\includegraphics[width=0.5\textwidth]{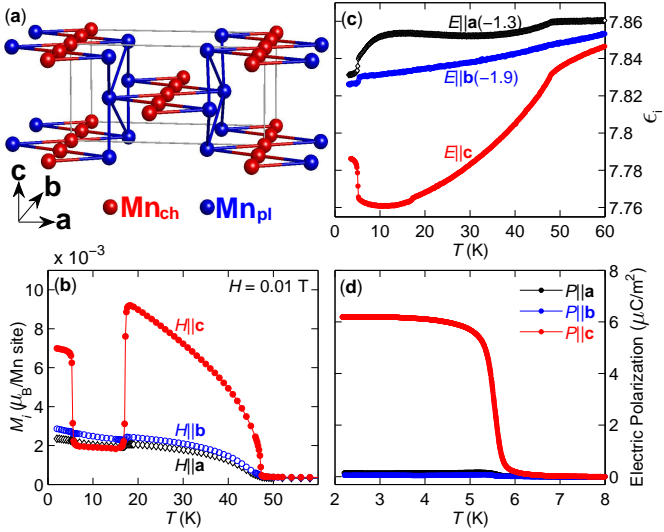}
\caption{(Color online) (a) The orthorhombic ($Pnma$) magnetic unit cell of Mn$_{2}$GeO$_{4}$. Mn$^{2+}$ ions occupy two sites; Mn$_{\rm ch}$ ions (red) are at the 4($a$) site, Mn$_{\rm pl}$ ions (blue) are at the 4($c$) site. Both the O$^{2-}$ anions and Ge$^{4+}$ cations are omitted for clarity. In the other panels we show the $T$-dependence of (b) the magnetization, $M_{i}$ (at 0.01~T, field-cooled), (c) the relative dielectric constant $\epsilon_{i}$ (at 0T), and (d) the electric polarization $P_{i}$ (at 0T), measured along all three principal axes.}
\label{fig:bulk_measurements}
\end{figure}

Mn$_{2}$GeO$_{4}$ crystallizes in the orthorhombic $Pnma$ (No.~62) space group~\citep{Cre70}. The $S=5/2$~\citep{Hag00} Mn$^{2+}$ ions occupy two distinct crystallographic sites octahedrally coordinated by O$^{2-}$ anions; the 4($a$) inversion site, which we call the chain Mn site, Mn$_{\rm ch}$ because they form chains along the \textbf{b}-axis, and the 4($c$) mirror site, which we call the plane Mn site, Mn$_{\rm pl}$, because they form distorted square lattice planes in the \textbf{b-c} plane [Fig.~\ref{fig:bulk_measurements}~(a)]. Direct nearest-neighbor (NN) exchange is anticipated to be small and the largest superexchange interaction is expected between Mn$_{\rm pl}$ ions in the \textbf{b-c} plane, where the angle between Mn$_{\rm pl}$-O-Mn$_{\rm pl}$ of NN Mn$_{\rm pl}$ ions is $124.43^{\rm o}$. Smaller interactions are expected within the Mn$_{\rm ch}$/Mn$_{\rm pl}$ zig-zag chains along the \textbf{b}-axis. The angle formed by the Mn$_{\rm ch}$-O-M$_{\rm ch}$ ions along the \textbf{b}-axis is $92.68^{\rm o}$, and the angle formed by the Mn$_{\rm ch}$-O-Mn$_{\rm pl}$ in the \textbf{a-b}-plane is $93.59^{\rm o}$, suggesting that these interactions are either weakly antiferromagnetic (AFM), or even FM. These Mn-O-Mn exchange geometries point towards both three-dimensional and frustrated interactions~\citep{Inter}, the latter being an almost ubiquitous feature of magnetic MFs~\citep{Che07}.\par

Large single crystals of Mn$_2$GeO$_4$ have been successfully grown by the floating zone method. The bulk properties of $M$, $P$, and the dielectric constant $\epsilon$ were measured along each crystallographic direction. Neutron diffraction (ND) experiments were performed at the Swiss Neutron Spallation Source SINQ, Paul Scherrer Insitut, Switzerland. Unpolarized ND was carried out using the TriCS and RITA-II instruments, while polarized ND (zero field neutron polarimetry) made use of the MuPAD device installed on the TASP instrument. Powder ND was also carried out using the diffractometer HRPT. Full details of our experiments can be found in Ref.~\onlinecite{Sup}.\par

The low temperature ($T$) bulk properties of Mn$_{2}$GeO$_{4}$ are summarized by the measurements shown in Figs.~\ref{fig:bulk_measurements}~(b)-~\ref{fig:bulk_measurements}(d). The $T$-dependence of the magnetization shows that long-range magnetic order sets in $T_{\rm N1}=47$~K, and that there are additional first-order transitions in the magnetic properties at both $T_{\rm N2}=17$~K and $T_{\rm N3}=5.5$~K. We label the phases as high-$T$ (HT), medium-$T$ (MT) and low-$T$ (LT) phases. These transitions are most clearly seen in the $T$-dependence of the FM moment that points along the crystal \textbf{c}-axis. Measurements of the relative dielectric constant along all crystal directions [Fig.~\ref{fig:bulk_measurements}(c)] show features that are clearly correlated with the magnetic transitions. The most pronounced anomalies are observed on cooling through $T_{\rm N3}$, with the largest response observed for $\epsilon_{c}$~$\left(E\parallel\textbf{c}\right)$. These data indicate the transition at $T_{\rm N3}$ to separate ferroelectric (FE) and paraelectric phases. Pyroelectric measurements [Fig.~\ref{fig:bulk_measurements}(d)] directly confirm the emergence of a macroscopic FE polarization $P$ parallel to the crystal \textbf{c}-axis for $T<T_{\rm N3}$. These data demonstrate that Mn$_{2}$GeO$_{4}$ features a ferromagnetic and ferroelectric phase.\par

\begin{figure}
\includegraphics[width=0.5\textwidth]{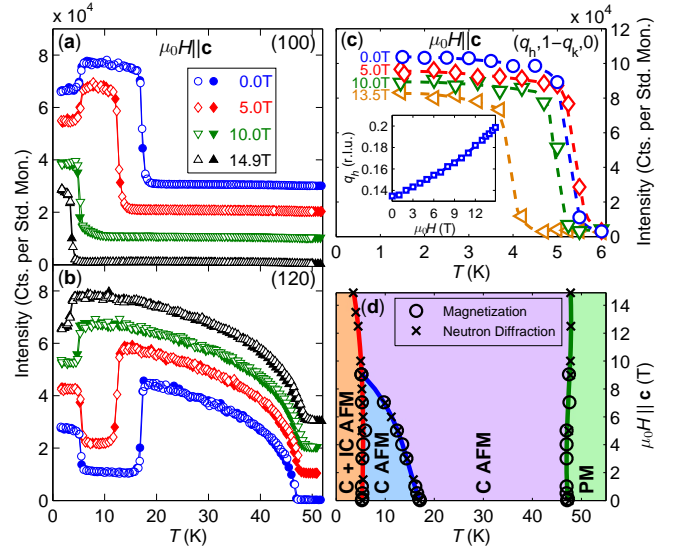}
\caption{(Color online) The $\mu_{0}H$- and $T$-dependence of the neutron intensity at the commensurate (C) (a) (100) and (b) (120) positions for fields $\mu_{0}H\parallel\rm {\textbf{c}}$. For clarity, in each of (a) and (b) adjacent field data are displaced vertically by 10000 counts per standard monitor. In (c) we show the $T$-dependence of the neutron intensity at the incommensurate (IC) position ($q_{h}$,1-$q_{k}$,0) for fields $\mu_{0}H\parallel\textrm{\textbf{c}}$. The inset to (c) shows the $\mu_{0}H$-dependence at $T=1.7$~K of the $q_{h}$ component of the incommensuration. In (d) we show ($\mu_{0}H$,$T$) magnetic phase diagram for $\mu_{0}H\parallel\rm {\textbf{c}}$ determined by both magnetization (data not shown) and neutron diffraction. Paramagnetic (PM), C AFM and C + IC AFM structure phases are labeled. In panels (a)-(c), open (filled) symbols correspond to $T$-warming (cooling) data and dashed lines correspond to guides for the eye. For all panels, error bars are of order the size of the data symbol.}
\label{fig:multi_panel_fig}
\end{figure}

Neutron diffraction directly identifies the nature of the magnetic order in the different phases. The $\mu_{0}H$- and $T$-dependence of the neutron Bragg intensity at the (100) and (120) positions, shown in Figs.~\ref{fig:multi_panel_fig}~(a-b), reveals commensurate (C) magnetic order with $\textrm{Q}_{\rm c}=(0,0,0)$ in all three magnetic phases, separated by first-order transitions. Furthermore, Fig.~\ref{fig:multi_panel_fig}~(c) shows that for $T<T_{\rm N3}$ long-range incommensurate (IC) order co-exists with the C order. The IC propagation vector is $\textrm{Q}_{\rm ic}=(q_{h},q_{k},0)$ where in zero field, $q_{h}=0.136(2)$ and $q_{k}=0.211(2)$. Fig.~\ref{fig:multi_panel_fig}(d) shows the ($\mu_{0}H\parallel\rm {\bf {c}}$,~$T$) magnetic phase diagram constructed from both $M$ and ND measurements. The MT phase becomes suppressed with increasing $\mu_{0}H$, and entirely absent by 9~T. The LT phase with co-existing C and IC order is more robust, with a noticeable suppression in $T_{\rm N3}$ only for $\mu_{0}H>$10~T, and the persistence of the IC order up to the highest fields applied in our experiments ($\mu_{0}H$=14.9~T). The inset of Fig.~\ref{fig:multi_panel_fig}~(c) shows the $q_{h}$ component of the IC wave-vector to increase monotonically with $\mu_{0}H$, in contrast to the $q_{k}$ component which remains $\mu_{0}H$-independent (data not shown).\par

\begin{table}
\caption{\label{tab:symmetry}Selected standard irreducible representations for the little group defined by the $Pnma$ spacegroup and $\textrm{Q}_{\rm c}=(0,0,0)$. $1$ is the identity, ${\mathcal{I}}$ is inversion about the origin, 2$_{\alpha}$ corresponds to a twofold rotation (or screw) axis, while $m_{\alpha,\beta}$ is a (mirror) or glide plane containing the axes $\alpha$ and $\beta$. The full character table is shown in Ref.~\citep{Sup}.}
\begin{ruledtabular}
\begin{tabular}{c|cccccccc}
 & 1 & 2$_{c}$ & 2$_{b}$ & 2$_{a}$ & ${\mathcal{I}}$ & $m_{ab}$ & $m_{ac}$ & $m_{bc}$\\
\hline\\
$\Gamma^{1}_{\rm c}$ & 1 & 1 & 1 & 1 & 1 & 1 & 1 & 1 \\
$\Gamma^{3}_{\rm c}$ & 1 & 1 & -1 & -1 & 1 & 1 & -1 & -1  \\
\end{tabular}
\end{ruledtabular}
\end{table}

\begin{table}
\caption{\label{tab:symmetry_IC} Both the standard irreducible representation ($\Gamma^{x}_{\rm ic}$) and the irreducible corepresentation ($D^{x}_{\rm ic}$) for the space group $Pnma$, and the general incommensurate propagation vector $\textrm{Q}_{\rm ic}=(q_{h},q_{k},0)$. Antilinear operators include the complex conjugation operator, $K$, and enter into the irreducible corepresentation only. $a=\textrm{exp}(-2\pi i q_{h}/2)$.}
\begin{ruledtabular}
\begin{tabular}{c|cccc}
                          & 1 & $m_{ab}$ & $K{\mathcal{I}}$ & $K2_{c}$ \\
\hline\\
$\Gamma^{1}_{\rm ic}/D^{1}_{\rm ic}$ & 1 &       $a$ &       1 &       $a$\\
$\Gamma^{2}_{\rm ic}/D^{2}_{\rm ic}$ & 1 &      -$a$ &       1 &      -$a$\\
\end{tabular}
\end{ruledtabular}
\end{table}

\begin{figure}
\includegraphics[width=0.48\textwidth]{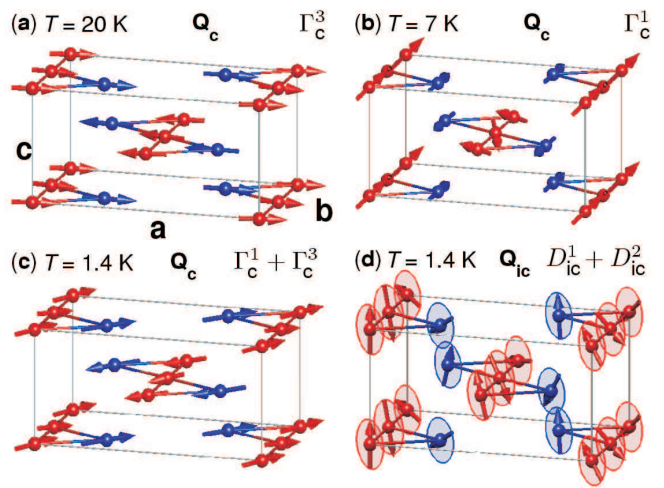}
\caption{(Color online) The refined zero field magnetic structures of Mn$_{2}$GeO$_{4}$. In (a)-(c) we show the refined commensurate structures at $T=20$~K, (b) $T=7$~K and (c) $T=1.4$~K. In (d) we show the refined incommensurate magnetic structure at $T=1.4$~K. In (d) the envelopes around the moments indicate the rotation plane of each ion.}
\label{fig:mag_structures}
\end{figure}

Using both unpolarized and polarized ND, and a magnetic symmetry analysis~\citep{Sup}, we find that the magnetic structures in the HT and MT phases are described by the standard irreducible representations (irreps) $\Gamma^{3}_{\rm c}$ [Fig.~\ref{fig:mag_structures}~(a)] and $\Gamma^{1}_{\rm c}$ [Fig.~\ref{fig:mag_structures}~(b)] defined in Table~\ref{tab:symmetry}, respectively~\citep{Sup}. In contrast to the single irrep descriptions of both the MT and HT phases, the C structure in the LT phase is best described by the irrep sum $\Gamma^{1}_{\rm c}+\Gamma^{3}_{\rm c}$ shown in Fig.~\ref{fig:mag_structures}~(c). Table~\ref{tab:symmetry} shows that the C magnetic structures allow for a FM magnetization in the HT and LT phases where $\Gamma^{3}_{\rm c}$ present, but not in the MT phase where only $\Gamma^{1}_{\rm c}$ ordered, consistent with the magnetization measurements. Since our ND data are insensitive to the small size of this FM moment ($<$0.01~$\mu_{B}$), the $m^{c}$ component on each site was constrained to be 0 in the refinements~\citep{Sup}.\par

Following the approaches in Refs.~\citep{Sch05,Rad07,Rad09}, a corepresentation (corep) analysis has been employed in order to determine the possible models for the IC magnetic structure in the LT phase - see~\citep{Sup} for details. The symmetry properties of both the irreducible corepresentation and the standard irreducible representation are reported in Table~\ref{tab:symmetry_IC}. There are just two unique coreps; each are of type `a', and each is generated by one of the standard irreps. Our refinements show that the IC magnetic structure is described by the corep sum $D^{1}_{\rm ic}+D^{2}_{\rm ic}$, though our experiments are insensitive to the relative phase between the corep mode amplitudes at each site. Later we will see symmetry to dictate that the mode amplitudes sum in quadrature so that the refined magnetic structure corresponds to slightly elliptical spin spirals on each Mn site as shown in Fig.~\ref{fig:mag_structures}~(d)~\citep{Sup}.\par

Our experimental data provide convincing evidence for a \emph{co-existence} of the C and IC orders in the FE phase, as opposed to a scenario where they are phase separated. The $\mu_{0}H$- and $T$-dependence of the neutron intensity for both the C and IC reflections [Figs.~\ref{fig:multi_panel_fig}~(a)~-~(c), and Ref.~\citep{Sup}] show that at all fields, just a \emph{single} transition $T$ separates \emph{both} C and IC orders in the FE phase from the C order in the other phases. Furthermore, both powder and single crystal measurements yield, within the experimental uncertainty, identical values for the ordered moments of the C and IC magnetic structures in the FE phase. We also find~\citep{Sup} that the maximum moment size for the C structure appears on sublattice Mn$_{\rm pl}$, while for the IC structure it appears on sublattice Mn$_{\rm ch}$, and that the total moment at each site arising from a superposed C and IC structure is always $\leq$~5~$\mu_{B}$, as expected for the free ion moment in the $S=5/2$ state of Mn$^{2+}$. \par

Inspecting the sequence of the magnetic order in Mn$_{2}$GeO$_{4}$, we identify two main motifs: 1) FM double-triangles in the $ab$-plane coupled antiferromagnetically, and 2) AFM non-collinear double-triangles. The stabilization of the FM order on the double-triangle may be possible because all its NN interactions are mediated by Mn-O-Mn bonds very close to $90^{o}$. Lowering the temperature, the FM double-triangles appear to become unstable due to smaller, presumably AFM interactions, and the magnetic order becomes non-collinear. Finally, in the FE LT phase, both FM and AFM tendencies are satisfied simultaneously at different wave-vectors.\par

We now show that the spontaneous $P$ along the \textbf{c}-axis is permissible according to the symmetry properties of the IC magnetic structure. By summing the corep mode amplitudes in quadrature as, for the example of magnetic site $j$, $\textbf{m}_{j}(D^{1}_{\rm ic})+i\textbf{m}_{j}(D^{2}_{\rm ic})$, we ensure that the point-group symmetry of the IC magnetic order is compatible with both the emergence, and the direction of the electric $P$. This can be seen by inspecting Table~\ref{tab:symmetry_IC}; since both $\mathcal{I}$ and $2_{c}$ enter the magnetic little group as antilinear operators, the signs of these operators invert for the corep with mode amplitudes preceded by the imaginary unit $i$~\citep{Rad07}. The resulting spin spiral preserves only the twofold rotation around the \textbf{c}-axis, and no operations for the \textbf{a}- or \textbf{b}-axes. This yields a point-group symmetry for the spiral of $\cdot\cdot2$ which is compatible with a $P$ observed \emph{only} along the \textbf{c}-axis [Fig~\ref{fig:bulk_measurements}~(d)]. In contrast, the point-group symmetries of alternative corep mode combinations that may explain the experimental data are incompatible with the observed $P$~\citep{Rad07}. Moreover, the symmetry properties of the C structure in the LT phase also do not permit an electric $P$ - strong evidence that the $P$ is associated solely with the onset of the IC spiral magnetic order. Precisely the same conclusions are reached by interpreting the refined symmetry properties of the LT magnetic structures in terms of the Landau theory for which the ME coupling is described by a trilinear interaction term~\citep{Ken05,Law05,Har07}.\par

\begin{figure}
\includegraphics[width=0.4\textwidth]{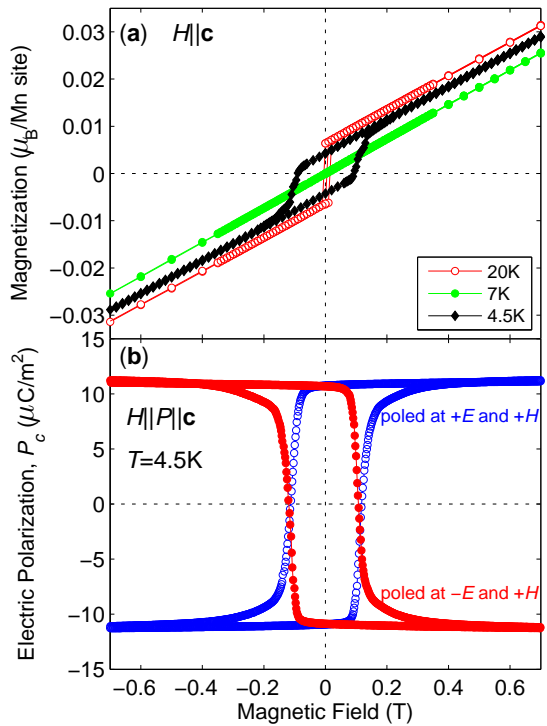}
\caption{(Color online) (a) The magnetic field dependence of the magnetization for $\mu_{0}H\parallel\textbf{c}$, and at selected temperatures. In (b) we show the magnetic field dependence of the electric polarization on poling the electric field.}
\label{fig:hysteresis_curves}
\end{figure}

$M$-$\mu_{0}H$ loops measured at temperatures within each magnetic phase [Fig~\ref{fig:hysteresis_curves}~(a)] show the expected hysteretic behavior in $M$ due to FM domain switching is only observed at $T=20$~K and $T=4.5$~K, which is where $\Gamma^{3}_{\rm c}$ is active. At $T=4.5$~K in particular, the coercive magnetic field corresponds well to that seen in the $P_{c}$-$\mu_{0}H$ hysteresis loops [Fig.~\ref{fig:hysteresis_curves}~(b)] demonstrating the synchronized switching of both FM and FE domains. This behavior suggests that the FM and FE domain walls are clamped in the MF phase, and that the directions of $M$ and $P$ remain fixed relative to one another across the MF domain.\par

We now show that an additional DM interaction in the FE phase is responsible for both the direction of the polarization and the magnetization, explaining their directly coupled nature. Because of the presence of two types of Mn sites, DM exchange interactions are allowed between all Mn$_{\rm ch}$ and Mn$_{\rm pl}$ neighbors, leading to the FM magnetization in the LT and HT phases. In contrast, the DM vector between nearest Mn$_{\rm ch}$ neighbors reads $(D_a, 0, D_c)$ in the paraelectric phase with $D_b=0$~\citep{Mor60}. The FE symmetry lowering makes $D_b \neq 0$, and thus provides an exchange interaction for the stabilization of the $m^{a}$ and $m^{c}$ components of the spin spiral. Further, $D_b \neq 0$ also provides a direct coupling between the $m^{c}$ component of $\Gamma^{1}_{\rm c}$ and the $m^{a}$ component of $\Gamma^{3}_{\rm c}$, which fixes the direction of the magnetization by removing a degree of freedom that would otherwise lead to additional possible magnetic domains. The sign of $D_b$ controls the sense of rotation of the spin spiral, and also the direction of the FM canting, thus providing the possibility for a direct coupling of ferromagnetism and ferroelectricity.

To summarize, our experiments have revealed Mn$_{2}$GeO$_{4}$ to be a new ferromagnetic ferroelectric. The direct coupling between ferromagnetism and ferroelectricity is mediated by a magnetic order composed of both incommensurate spiral and canted commensurate spin structure components. Our study shows multi-component magnetic structures as a new viable route for the design of functional materials that allow the precision control of multiple orders.

We thank V.~Pomjakushin for assistance with the neutron powder diffraction measurements. We acknowledge financial support from the Swiss National Centre of Competence in Research program MaNEP, KAKENHI (20674005, 20001004), MEXT, Japan, and the SNF under grant number 200021$_{-}$126687.

\bibliography{mn2geo4_bib_2}

\end{document}